\newif\iffullversion
\fullversiontrue

\iffullversion  
 \documentclass[11pt,letterpaper]{article}
\usepackage[letterpaper,margin=1in]{geometry}
\else
\documentclass[sigconf, anonymous]{acmart}
\setcopyright{none} 
\acmConference[Anonymous Submission to ACM CCS 2019]{ACM Conference on Computer and Communications Security}{Due 15 May 2019}{London, TBD}
\acmYear{2019}
\fi
\usepackage[bookmarks=false]{hyperref}
\usepackage{xspace,amsmath,amsfonts,amssymb,tikz,multirow,float,newclude,pgfplots,bm, enumitem}
\usepackage{xspace,amsmath,amsfonts,tikz,multirow, xcolor}
\usepackage{array}
\usepackage{xcolor,colortbl,tikz}
\usepackage{lipsum}
\usepackage{multicol}
\usepackage{cite}
\usepackage{lipsum}
\usepackage{multicol}
\usepackage{changepage}
\usepackage[T1]{fontenc}
\usepackage{amsmath}
\usepackage{amssymb}
\usepackage{amsthm}

\usetikzlibrary{matrix,shapes,arrows,positioning,chains, calc,backgrounds}

\newcommand{\adv}{\ensuremath{\mathcal{A}}\xspace}

\newcommand{\namedref}[2]{\hyperref[#2]{#1~\ref*{#2}}}

\newcommand{\sectionref}[1]{\namedref{Section}{#1}}

\newcommand{\figureref}[1]{\namedref{Figure}{#1}}
\newcommand{\subfigureref}[2]{\hyperref[#1]{Figure~\ref*{#1}#2}}

\definecolor{darkred}{rgb}{0.5, 0, 0} 
\definecolor{darkgreen}{rgb}{0, 0.5, 0} 
\definecolor{darkblue}{rgb}{0,0,0.5} 

\pdfoutput=1

\hypersetup{
    colorlinks=true,
    linkcolor=darkred,
    citecolor=darkgreen,
    urlcolor=darkblue   
}

\newcommand{\todo}[1]{%
    \mbox{}
    \marginpar{%
        \colorbox{red!80!black}{\textcolor{white}{to-do}}%
        \vspace*{-22pt}
    }%
    \textcolor{red}{#1}%
}

\newcommand{\hgen}[1]{\textsf{HGen}{#1}}
\newcommand{\henc}[1]{\textsf{HEnc}{#1}}
\newcommand{\hdec}[1]{\textsf{HDec}{#1}}
\newcommand{\hsum}[1]{\textsf{HSum}{#1}}
\newcommand{\hmul}[1]{\textsf{HMul}{#1}}
\newcommand{\G}{G\xspace}
\newcommand{\Z}{\mathbb{Z}}

\newcommand{\sT}{\mathbf{T}}

\newcommand{\Fgc}{\mathcal{GC}}

\newcommand{\remove}[1]{}
\newcommand{\maliciousremove}[1]{}
\newcommand{\pirge}{\textsf{PIR.Gen}}
\newcommand{\pirqe}{\textsf{PIR.Query}}
\newcommand{\pirexpand}{\textsf{PIR.Expand}}
\newcommand{\pirans}{\textsf{PIR.Answer}}
\newcommand{\pirextract}{\textsf{PIR.Extract}}

\newcommand{\dect}{\textsf{Epione}\xspace}
\newcommand{\tpsi}{$t$-PSI\xspace}
\newcommand{\psica}{PSI-CA\xspace}
\newcommand{\fpsi}{$f$-PSI\xspace}

\newcommand{\ctgenerate}{\texttt{Generate}}
\newcommand{\ctexchange}{\texttt{Exchange}}

\newcommand{\ctquery}{\texttt{Query}}

\definecolor{highlightcolor}{HTML}{F5F5A4}
\definecolor{highlighttextcolor}{HTML}{000000}

\newcommand{\mathhighlight}[1]{\basehighlight{$#1$}}

\newcommand{\basehighlight}[1]{\colorbox{highlightcolor}{\color{highlighttextcolor}#1}}

\def\Sim{\ensuremath{\mathsf{Sim}}}

\newcommand{\OUT}{\textsf{OUT}\xspace}
\newcommand{\real}{\textsf{REAL}\xspace}
\newcommand{\ideal}{\textsf{IDEAL}\xspace}
\newcommand{\Ss}{\ensuremath{\mathcal{S}}\xspace}

\newcommand\mydef{\mathrel{\overset{\makebox[0pt]{\mbox{\normalfont\tiny\sffamily def}}}{=}}}

\ifdefined \qed
\relax
\else
\newcommand{\qed}{\hspace*{\fill}\rule{7pt}{7pt}}
\fi

\ifdefined \theorem
\relax
\else
\newtheorem{theorem}{Theorem}
\newtheorem{definition}{Definition}

\newenvironment{proof_sketch}{\quad\par\noindent{\bf Sketch of proof:~~}}{\qed\quad}

\fi

\renewcommand\to{\ensuremath{\rightarrow}}
\newcommand\from{\ensuremath{\leftarrow}}

\newcommand{\ceil}[1]{{\lceil{#1}\rceil}}

\newcommand{\set}[1]{\ensuremath{\{#1\}}}

\def\bool{\set{0,1}}

\newcommand{\abs}[1]{\lvert#1\rvert}

\newlength{\protowidth}

\newlength{\saveparindent}
\newlength{\saveparskip}
\newcounter{ctr}

\newcounter{ectr}

\def\numinst{\ensuremath{m}}

\def\rm{\ensuremath{r_\numinst}}

\def\aa{\ensuremath{\alpha}}

\usepackage{mathtools}

\def\bu_i{\ensuremath{\boldsymbol{u_i}}}
\def\bu{\ensuremath{\boldsymbol{u}}}
\def\bct_i{\ensuremath{\boldsymbol{ct_i}}}

\def\bct{\ensuremath{\boldsymbol{\tilde{t}}}}

\begin{document}

\iffullversion

\title{
  \dect: Lightweight  Contact Tracing with Strong Privacy
}
\author{
	Ni Trieu\thanks{UC Berkeley, \textsf{nitrieu@berkeley.edu}}
	\and Kareem Shehata\thanks{kareem@shehata.ca}	
	\and Prateek Saxena\thanks{National University of Singapore, \textsf{$\{$prateeks, reza$\}$@comp.nus.edu.sg}}
	\and Reza Shokri\footnotemark[3] 
	\and Dawn Song\footnotemark\thanks{UC Berkeley and Oasis Labs, \textsf{dawnsong@berkeley.edu}}	
}
\date{\vspace{-5ex}}

\else
\title{
\dect: Decentralized Contact Tracing with Strong Privacy
}

%
\fi

\iffullversion
\maketitle
\fi

\begin{abstract}

    Contact tracing is an essential tool in containing infectious diseases such as COVID-19. Many countries and research groups have launched or announced mobile apps to facilitate contact tracing by recording contacts between users with some privacy considerations.  Most of the focus has been on using random tokens, which are exchanged during encounters and stored locally on users' phones. Prior systems allow users to search over released tokens in order to learn if they have recently been in the proximity of a user that has since been diagnosed with the disease.  However, prior approaches do not provide end-to-end privacy in the collection and querying of tokens. In particular, these approaches are vulnerable to either linkage attacks by users using token metadata, linkage attacks by the server, or false reporting by users.
    
    In this work, we introduce \dect, a lightweight system for contact tracing with strong privacy protections. \dect alerts users directly if any of their contacts have been diagnosed with the disease, while protecting the privacy of users' contacts from both central services and other users, and provides protection against false reporting. As a key building block, we present a new cryptographic tool for secure two-party private set intersection cardinality (PSI-CA), which allows two parties, each holding a set of items, to learn the intersection size of two private sets  without revealing intersection items. We specifically tailor it to the case of large-scale contact tracing where clients have small input sets and the server's database of tokens is much larger.

\end{abstract}

\iffullversion
\else
\maketitle
\fi

\section{Introduction}
\label{sec:intro}

Contact tracing is an important method to curtail the spread of
infectious diseases. The goal of contact tracing is to identify
individuals that may have come into contact with a person that has been diagnosed with the disease, so they can be isolated and tested, and thus prevent the spread of the disease any further.

In the ongoing COVID-19 pandemic, recording of individuals in close proximity has been
facilitated by mobile apps that detect nearby mobile phones using Bluetooth. Several countries have been developing contact tracing apps. Such large-scale collection of personal contact
information is a significant concern for privacy~\cite{PrivateKit,cho2020contact}.

The main purpose of contact tracing applications---recording the
fact that two or more individuals were near each other at a certain
moment of time---seems to be at odds with their privacy. The
app must record information about the individual's personal contacts
and should be able to reveal this information (possibly, on demand) to some
authorities. Indeed, in a fully untrusted environment one should expect any
participant to behave adversarially with the goal to exploit others'
personal information. Further, both end users of tracing
apps as well as the authorities using the collected data can become victims of security attacks, which can allow powerful
adversaries to misuse the information collected by the app.

Multiple ways have been proposed to protect the users' private proximity data,
offering different privacy guarantees and coming at different
implementation costs. For instance, in the recently released BlueTrace protocol used by the Singapore Government~\cite{ttg}, users are guaranteed privacy from each other, but this model places complete trust in certain operating entities for protecting user information. We analyze the BlueTrace proposal and other approaches in more detail in Section~\ref{subs:rel_work_ct}.

We consider a model where the government authorities do not store any identifying user information (e.g. phone numbers and social contacts). Databases of such sensitive information are lucrative targets for cyber-attackers, and in many countries the collection of contact information conflicts with privacy
regulations and public concerns, which may create a barrier to the usage of the tracing service. This is important, as privacy concerns may hinder adoption in some jurisdictions and contact tracing is expected to be effective only when participation is high (e.g. 60\% or more of the population~\cite{Ferretti2020}). 

In our model, the health authorities maintain a database of tokens corresponding to users which have been diagnosed with the disease. The user's tracing app periodically checks an untrusted server to check if the user is potentially at risk in such a way that the server cannot deduce any information about the user which is not implied by the desired functionality and the user learns no information beyond whether they may have been exposed to the disease.

Our model can also be contrasted to several other decentralized mobile contact tracing system/protocols, which will also be analyzed in Section~\ref{subs:rel_work_ct}. As we will see through that analysis, existing proposals or launched systems are vulnerable to one or more of the following privacy attacks:

\begin{itemize}
	\item[(1)] {\em Infection status / exposure source by users}: If tokens of users diagnosed positive are publicly released, Alice can determine which publicly-posted tokens matches the log on her phone. This could reveal the time, for example, when Alice and the user diagnosed positive with the disease were in close proximity, enabling her to identify Bob. Such identification is undesirable as people have been reported to harass individuals suspected to be the source of exposure to the disease~\cite{korea-harassment}, leading to the so-called "vigilante" problem.
	
	\item[(2)] {\em Infection status by server}: If the server can determine which users have been diagnosed with the disease, this leaks the infection status of users to the server operator. This may not be a concern in jurisdictions where the server is operated by the health authority which already knows this information. However, in jurisdictions where the server is operated by another party that does not or should not have this information, this form of linkage can be a serious privacy threat.
	
	\item[(3)] {\em Social graph exposure and user tracking}: If a central database is used to collect both sent and received tokens as in \cite{Covid-watch}, or it is possible to infer the source of a sent token as in the case of \cite{TraceTogether}, then the operator of this server is able to deduce all of the social connections of a user that is reported positive for the disease, including when and for how long each contact was made.
	
	This linkage can also be used to track a user's movements via Bluetooth beacons. Bluetooth has protections against tracking users over time introduced in Bluetooth 4.0 Low Energy \cite{bluetooth-privacy}. With solutions such as \cite{google-apple,chan2020pact} it is possible to link previously seen Bluetooth device (MAC) addresses despite these protections via exchanged contact tokens once the seeds used to generate the tokens are released\footnote{As internet users have already pointed out https://twitter.com/moxie/status/1248707315626201088}, producing a similar attack to \cite{bluetooth-tracking}.
	
	\item[(4)] {\em False claims by users}: Users may falsely claim to be diagnosed positive or may claim to be in contact with someone by modifying their app logs. 
	For example, a user who has recovered after being diagnosed positive may threaten to retroactively add a user not currently diagnosed with the disease to their contact log, unless they are paid a ransom. Similarly, a user not diagnosed may falsely claim to be in contact with a user diagnosed positive whose tokens have been revealed, for example, to qualify for a diagnostic test. 
	
\end{itemize}

Table \ref{tbl:comp-complx} provides a brief comparison of different contact tracing systems with respect to different security/privacy properties, required computational infrastructure and client's communication cost, all of which are important for real-world contact tracing. Details of the systems compared will be discussed in Section~\ref{subs:rel_work_ct}.

\begin{table*}[]
\footnotesize
	\begin{adjustwidth}{-1cm}{-1cm}
	\centering
	\begin{tabular}{|l||c|c|c|c|c|c|c|}
		\hline
		\multicolumn{1}{|c|}{\multirow{3}{*}{\textbf{System}}}                        & \multicolumn{2}{c|}{\textbf{System Req.}}                 & \multicolumn{4}{c|}{\textbf{Privacy Protection Against}}                      & \textbf{Client}             \\ 
		\multicolumn{1}{|c|}{}                                               & Trusted                      & \# & \multicolumn{2}{c|}{Infection Status} & Social & False-positive       & Comm.                \\ 
		\multicolumn{1}{|c|}{}                                      & \multicolumn{2}{c|}{Server}                                               & By User          & By Server  & Graph           & User          & Cost                   \\ \hline \hline
		TraceTogether~\cite{ttg}                                                        & Yes                          & 1                         & Yes              & No & No              & Yes                       & $O(n)$     \\ \cline{1-8}
		Baseline*                 & \multirow{3}{*}{\textbf{No}} & 1                         & No               & No     & Most          & Some                      & \multirow{2}{*}{$O(N)$} \\ \cline{1-1} \cline{3-7} 

	Private Messaging~\cite{cho2020contact} &                              & 3                         & No               & Yes   & Yes         & No                      &                    \\ \cline{1-1} \cline{3-7}\cline{8-8}

	\dect                                                 &                              & 2                      &   \textbf{Yes}    & \textbf{Yes} & \textbf{Yes}   & \textbf{Yes}   &  $O(n\log(N))$   \\ \hline
	\end{tabular}
\end{adjustwidth}

\caption{Comparison of contact tracing systems with respect to security, privacy, required computational infrastructure, and client communication cost. \textbf{Baseline} systems include Private Kit\cite{PrivateKit}, Covid-watch~\cite{Covid-watch}, CEN~\cite{CEN}, DP-3~\cite{DP-3T}, and PACT's baseline system~\cite{chan2020pact}. Some of these systems provide a limited level of false-positive claim protection with an additional server (or healthcare provider), and most provide protection from social graph discovery.  $N$ is the total number of contact tokens from users diagnosed positive with the disease, $n$ is the number of contact tokens recorded by an average user that need to be checked for disease exposure (Note that $\frac{N}{n}$ is typically the number of new positive diagnoses per day, thus $n<<N$).}

\label{tbl:comp-complx}
\end{table*}

\subsection{Our Contribution}

In this work, we introduce \dect, a new system for decentralized contact tracing with stronger privacy protections than the strongest models currently found in related work. As a key primitive enabling \dect, we introduce a new private set intersection cardinality or \psica, which is used to check how many tokens held by a user (client) match the tokens in a set stored on a server, without the user revealing their tokens. More formally, \psica allows two parties, each holding a private set of tokens, to learn the size of the intersection between their sets without revealing any additional information. Our \psica primitive is designed to be efficient for a large server-side database and a small client-side database, as is the case for contact tracing applications. 
Our new \psica constructions allow us to meet all of our privacy goals.  With several other optimizations in our system design, we show that \psica can make privacy-preserving contact tracing practical.

In summary, we make the following contributions:
\begin{itemize}
	\item We design \dect, an efficient high-performance contact tracing system that provides strong privacy guarantees, specifically that user contact information is not revealed beyond what is desired to any party and that the diagnosis status is revealed only to health authorities. The system prevents all important attacks, such as linkage attacks (e.g. infection status, social graph exposure), and false-positive claims, to which current models underpinning related work are vulnerable.
	
	\item We propose a new semi-honest private set intersection cardinality (\psica) primitive for asymmetric sets. 
	Our \psica protocol has communication complexity linear in the size of the smaller set ($n$), and logarithmic in the larger set size $N$. More precisely, if the set sizes are $n << N$, we achieve a communication overhead of $O(n \log N)$. 
\end{itemize}

\subsection{System Overview}
\label{subs:system_overview}

\begin{figure*}[t]
	\centering
	\includegraphics[scale=0.65]{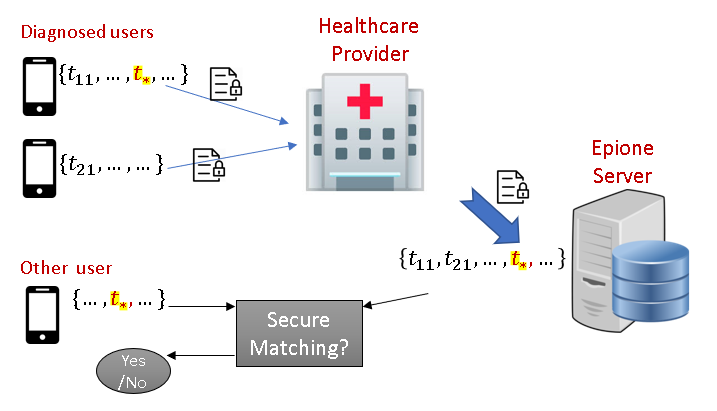}
	\caption{Overview of our \dect system. When a person is diagnosed with the disease, the healthcare provider collects their sent tokens and forwards them to the \dect server encrypted under the server's public key (actually, the PRG seed is collected to reduce communication costs). The \dect server decrypts the received ciphertexts from the healthcare provider and obtains the transmitted tokens of all patients diagnosed positive. Next, each user's app uses \psica to compare their set of received tokens with the set of sent tokens stored on the \dect server. If the intersection size is more than zero, then the user is alerted that they may have been exposed to the disease.}	
	\label{fig:overview_system}
\end{figure*}

\figureref{fig:overview_system} shows an overview of the \dect system. Users of \dect want to be notified if any of the people they have been in contact with are later diagnosed with the disease. They do \textit{not} want to reveal to other users their identity, reveal whether they have been diagnosed positive, be tracked over time, or reveal their contacts to any other organization.

We use a short-range network (such as Bluetooth) to detect when two users are within close range and exchange a randomly generated ``contact token''. All of the sent and received contact tokens are stored securely on the user's phone in the ``sent token list'' and ``received token list'', respectively. The received token list never leaves the user's phone in a form that can be used by anyone else, and the sent token list is only revealed to a healthcare provider on a positive diagnosis and with the user's consent. In \sectionref{sect:system}, we explain in detail how to generate and store the tokens.

In \dect, we assume that there is an untrusted service provider, which we call the \dect Server (or the backend server or just ``server'' for simplicity when it isn't ambiguous), which can collect the transmitted contact tokens from all users tested positive with the disease.
The \dect server allows users to check whether they have received a token from a user who has since been diagnosed with the disease, without revealing to the server their tokens (and thus their contacts) and without the server revealing any information to the user about the tokens of users diagnosed positive beyond the count of contact tokens in common. We use secure computation techniques, particularly \psica, for private matching. This prevents the \dect server from inferring linkages between users, as well as preventing users from inferring the diagnosis status of other users, or the source of any exposure to the disease.

It is assumed that a healthcare provider (such as a hospital)
is aware of the identity of the user whom it diagnoses as having the disease. Thus, exposing the identity of the user diagnosed positive from the provider is not considered as a threat. 
It is also assumed that the healthcare provider keeps a local database of positively diagnosed users, to be able to verify if a user was legitimately diagnosed positive. The healthcare provider collects (with the user's consent) the list of ``sent tokens'' from a positively diagnosed user's app and sends it to the \dect server, which the latter adds to a database of contact tokens from such users.

Note that in this model the server does not know the identity of the user diagnosed positive. It is not hard to imagine collusion between the healthcare database and the backend server for \dect, say by a state actor or attacker within the healthcare provider. Even then, the sent tokens are not useful for identifying any contacts or any other private information. Since tokens are randomly generated, the attacker would need to know which users received those tokens to re-identify them. We will show later that the \dect server never learns the received tokens of any user and thus linkage is not possible.

\section{Related Work} 

We begin by discussing previous approaches to contact tracing, and then current approaches to secure computation and private set intersection, which will form the basis for our own \psica used in \dect.

\subsection{Contact Tracing Approaches}
\label{subs:rel_work_ct}
Due to the rapid spread of the COVID-19 pandemic and the importance of contact tracing, many research groups have been developing tools to improve contact tracing. Most schemes either (1) rely on and expose data to a trusted third-party, such as TraceTogether \cite{ttg}, or (2) uses a decentralized/public list approach such as COVID-Watch~\cite{Covid-watch}, PACT~\cite{chan2020pact}, or Google/Apple~\cite{google-apple} that allows users to infer linkages such as exposure sources. We divide existing contact tracing systems into Trusted Central Authority and Untrusted Centralized / Distributed approaches. 

\subsubsection{Trusted Third-Party (Centralized) Approaches}
The primary example of a trusted third party approach is introduced in the TraceTogether app by the Singapore Government, released on March 20, 2020~\cite{TraceTogether}. Based on its underlying Bluetrace protocol specification, the protocol abstractly works as follows. Let Alice and Bob be users of the app, and let Grace be the government server (or other central authority).

\begin{enumerate}

\item\textbf{Setup.} Alice and Bob both install the app on their smartphones.
During the setup process, both Alice and Bob register a phone number
with Grace and are each assigned a
unique identifier (i.e., $ID_A$ for Alice, and
$ID_B$ for Bob). Grace stores the phone numbers and ID numbers registered in a database.

\item\textbf{Token Generation} Grace assigns to Alice and Bob a set of contact tokens\footnote{In Bluetrace, tokens are called "TempIDs". We have kept the term token here for comparison with other systems.} that are to be used at different times. Tokens are generated by concatenating the user's ID, the start time of the token, and the end time of the token, and then encrypting with AES-256-GCM with a key known only to Grace. Tokens are thus expressed as: $T_i = E_{AES}(k, ID || t_{start} || t_{end})$. Tokens are sent in batches along with validity times to users' devices in case they are not able to fetch new tokens on demand\footnote{Note that tokens are generated by Grace mainly to reduce computational load on client devices. This scheme is equivalent to having clients produce tokens by encrypting with a public key for Grace with the same information.}.

\item\textbf{Contact.}
When Alice and Bob meet, they exchange tokens valid for the current time window.
Both devices keep a list of received tokens.

\item\textbf{Diagnosis.} Some time later, Bob is confirmed to have
contracted the disease. Bob sends his received tokens list (i.e. the list that contains
$T_A$ received from Alice) to Grace. Grace
verifies that Bob does in fact have a positive test result (this can
be done either by matching Bob in a separate database of test results,
or by providing Bob with a certificate that validates his positive result).

\item\textbf{Follow-up.} Grace then decrypts each token received from Bob using her key. This reveals the user ID and time of each contact along with other metadata. Grace can then look up each user's (e.g. Alice's) phone number and directly follow up with them.

\end{enumerate}

Clearly, this model places trust in the government health authority (Grace). When Bob is diagnosed with the disease and submits his tokens, Grace learns all of Bob's contacts, including when and for how long they were in contact. If Grace's key is compromised by an attacker, it could be used to track all users via Bluetooth (similar to \cite{bluetooth-tracking}).
The Bluetrace whitepaper has explained the rationale for the proposed model and alluded to concerns with it.


\subsubsection{Untrusted Third-Party or Decentralized Approaches}

Nearly all of the proposed or launched contact tracing schemes that do not rely on a trusted authority use a scheme as follows, with minor variations:

\begin{enumerate}
    \item Alice and Bob, two users of the contact tracing scheme, download and install the app.
    \item Alice and Bob both generate contact tokens that rotate over time and cannot be used to reveal their identity directly or track them over time.
    \item When Alice and Bob meet, they exchange contact tokens, for example via Bluetooth\footnote{in GPS or geolocation based systems, Alice and Bob independently generate tokens as a function of location and time, for example by hashing a grid square and time quantum.}. Alice and Bob both keep a list of received tokens and sent tokens.
    \item Bob is later diagnosed with the disease. Bob submits his tokens to an untrusted server. This can be either the list of received tokens, the list of sent tokens, or both. Alternatively, Bob can submit the secret used to generate tokens from his sent list.
    \item A list of tokens from users diagnosed with the disease is then maintained either in a private database to which users can submit queries or published in a public list so that users can check for intersections on their device.
    \item If a user finds they have tokens in common with the public list, or via querying a central database, they may have been exposed to the disease and should be notified to seek appropriate next steps.
\end{enumerate}

Covid-watch~\cite{Covid-watch}, Private Kit~\cite{PrivateKit}\footnote{PrivateKit claims that in V3 they will introduce strong privacy protections, but as of writing this paper the protocol to do so has not been announced.}, PACT's baseline design~\cite{chan2020pact}, and Google/Apple~\cite{google-apple} are all variations on this design. Some use pseudo-random number generators, and upload seeds for the sent token lists to reduce communication and storage costs at the expense of greater cost for comparisons, but this has no impact on privacy implications. All of these designs are susceptible to linkage attack by either users, the server, or both. Some offer protection against false positive claims.

In all of the above systems, each phone has to compare the publicly posted contact tokens against their own history, which requires to them download all public tokens. This requires significant bandwidth and places a burden on mobile devices.

\subsubsection{Privacy Improvements}

\textbf{Private Messaging.} To reduce the linkage information learned by a central server, \cite{cho2020contact} proposes to use two or more non-colluding servers operating as a private messaging network between users and the central server operated by the government (Grace). Concretely, assume that Grace stores a collection of mailboxes, one for each token that Alice and Bob exchange, and there are two non-colluding communication servers Frank and Fred. Frank forwards messages to/from Fred, and Fred forwards messages to/from Grace, all in such a way that Grace cannot know the source or destination of any messages. When Alice and Bob are in close contract, they exchange tokens. At fixed time points, both parties send a message which contains their current diagnosis status to each other via Frank, Fred, and Grace. For example, Bob addresses the
message to Alice encrypted using Alice's public key, and gives the message to Grace (via Frank and Fred), who puts it in Alice's mailbox. Alice checks all of the mailboxes through Frank and Fred to learn whether she has been exposed to the virus. Since Alice sees Bob's message in her mailbox,  Alice might be able to infer who
Bob is based on the time they are nearby. Moreover, Grace needs to maintain all tokens (messages) of all users, which requires storage.

\textbf{Re-randomization of tokens.} An alternative approach presented in PACT~\cite{chan2020pact} aims to prevent linkage attacks by users by re-randomizing tokens. Their proposed solution is based on the DDH assumption, and works with the following changes compared to their baseline system:
\begin{itemize} [noitemsep]
	\item Alice generates a token in the form $T_a = (g^{r_i}, g^{r_i s_A})$, where $s_A$ is a key that Alice never shares, and $r_i$ is a nonce used only for this token
	\item When Alice and Bob meet, Alice gives $T_a$ to Bob
	\item When Bob is diagnosed positive, he chooses a random $r'$. If $T_a = (x, y)$, he submits the pair $T_a' = (x^{r'}, y^{r'})$ to the server
	\item Alice can determine whether she is at risk by checking all of the tokens in the public list to find one that satisfies $y=x^{s_A}$
\end{itemize}
 
While generating tokens is almost free in our \dect, PACT with token re-randomization requires two group elements for each token's generation. Moreover, as mentioned by the authors, the privacy benefit inherently relies on each user re-using the same secret key ($s_A$ in this case), and they cannot force a malicious user to comply. Using a different $s_A$ for different encounters allows Alice to determine which encounter caused her exposure to the disease. In contrast, this malicious action does not happen in our \dect.

\subsection{Secure Computation and Private Set Intersection}
Private set intersection (PSI) refers to a cryptographic protocol that allows two parties holding private datasets to compute the intersection of these sets without either party learning any additional information about the other's dataset. PSI has been motivated by many privacy-sensitive real-world applications. Consequently, there is a long list of publications on efficient secure PSI computation~\cite{NDSS:HuaEvaKat12,USENIX:PSSZ15, CCS:KKRT16, CCS:KMPRT17, CCS:CheLaiRin17, C:PRTY19}. However, PSI only allows the computation of the intersection itself. In many scenarios it is preferable to compute some function of the intersection rather than reveal the elements in the intersection, such as whether intersection size is more than a given threshold $f$, as in contact tracing. Limited work has focused on this so-called \fpsi problem. Further, most prior works assume sets of comparable size that can be communicated in their entirety to the other party.
In this section we focus on protocols~\cite{NDSS:HuaEvaKat12, USENIX:PSSZ15,EC:PSWW18, EC:PSTY19, CCS:CHLR18,EPRINT:IKNPRSSSY19} that support \fpsi as well as \psica.

\paragraph{GC-based \fpsi.}  Huang, Katz, and Evans~\cite{NDSS:HuaEvaKat12} presented techniques for using the generic garbled-circuit approach for \fpsi, which is based on their efficient sort-compare-shuffle circuit construction.  Later Pinkas et al~\cite{EC:PSWW18, EC:PSTY19} improved a circuit-PSI using several hashing techniques.

The main bottleneck in the existing circuit-based protocols is the number of string comparisons and computing the statistics (e.g, count) of the associated values that are done inside a generic circuit-based secure two-party computation, which is communication-expensive.

\paragraph{HE-based \fpsi.} Similar to GC-based \fpsi, the protocol of~\cite{CCS:CHLR18} uses HE-based PSI to return additive secret share of the common items, which can be forwarded as input to a secondary generic circuit-based protocol. While their protocol has the communication complexity logorithmic in the larger set size, it requires at least two more interactive rounds and a certain amount of cost to implement the second MPC protocol. Therefore, the current Diffie-Hellman Homomorphic encryption approach of~\cite{EPRINT:IKNPRSSSY19} is still preferable in many real-world applications\footnote{\url{https://security.googleblog.com/2019/06/helping-organizations-do-more-without-collecting-more-data.html}}, due to their reasonable communication complexity of \fpsi protocols. However, the protocol of~\cite{EPRINT:IKNPRSSSY19} has communication complexity linear in  set size, which is still communication-expensive in client-server settings.

\section{Problem Statement and Security Goal}
\label{sec:goals}

Here we define the problem of contact tracing that we intend to solve, and our security goals for \dect.

\subsection{Problem Definition}
\label{subs:probl}
We define the problem of contact tracing based on token exchange as follows. Various clients communicate with each other and with a contact tracing service. The service is provided by one or more servers. The overall system consists of the following procedures:
\begin{itemize} [noitemsep]
	\item $\ctgenerate(\kappa) \to t$: Client uses the $\ctgenerate$ function to generate contact tokens, $t$, to be exchanged with other users.  The function takes a security parameter $\kappa$ as input.
	
	\item $\ctexchange(t_a) \to t_b$: The client (client a) uses the \ctexchange\ function to exchange tokens with another user (client b). The client's token $t_a$ is sent to the other user, and the client receives $t_b$ from the other user. The client (client a) stores $t_b$ in the ``received tokens list''. Similarly, the other user (client b) stores $t_a$ in their ``received tokens list''.
	
	\item $\ctquery(T_R, S) \to a$: With a set $\sT_R$ of received tokens from \ctexchange, the client uses the \ctquery\ function to query the server $S$ and get answer a indicating how many of their tokens came from users diagnosed positive for the disease.
	
\end{itemize}


\subsection{Security Definition and Goal} 
\label{subs:secure_def}

We consider a set of parties who have agreed upon a single function $f$ to compute (such as contact tracing) and have also consented to give $f$'s final result to some particular party. At the end of the computation, nothing is revealed by the computational process except the final output. In the real-world execution, the parties often execute the protocol in the presence of an adversary \adv who corrupts a subset of the parties. In the ideal execution, the parties interact with a trusted party that evaluates the function $f$ in the presence of a simulator \Sim 
~that corrupts the same subset of parties. There are two classical security models.
\begin{itemize}[noitemsep]
	\item Colluding model: This is modeled by considering a single monolithic adversary that captures the possibility of collusion between the dishonest parties. The protocol is secure if the joint distribution of those views can be simulated.
	\item Non-colluding model: This is modeled by considering independent adversaries, each captures the view of each independent dishonest party. The protocol is secure if the individual distribution of each view can be simulated.
\end{itemize}

There are also two adversarial models.
\begin{itemize}[noitemsep]
\item Semi-honest model (or honest-but-curious): The adversary is assumed to follow the protocol, but attempts  to obtain extra information from the execution transcript.

\item Malicious model: The adversary may follow any arbitrary polynomial-time strategy to deviate from the protocol, such as supplying inconsistent inputs, or executing different computation.

\end{itemize}

For simplicity, we assume there is an authenticated secure channel between each pair of clients, and client-server pair (e.g., with TLS).  In this work, we consider a model with non-colluding servers. We formally present the security definition
considered in this work, which follows the definition of~\cite{Goldreich09, EPRINT:KamMohRiv12}.

\textbf{Real-world execution.} The real-world execution of protocol $\Pi$ takes place between users $(P_1,\ldots, P_n)$, servers $(P_{n+1},\ldots, P_N)$ and adversaries $(\adv_1,\ldots, \adv_m)$, where $m < N$. Let $H \subseteq [n]$ denote the honest parties, $I \subseteq  [n]$ denote the set of corrupted and non-colluding parties and $C \in [n]$ denote the set of corrupted and colluding parties.

At the beginning of the execution, each user $P_{i \in [n]}$ receives its input $x_i$, an auxiliary input $z_i$ and random tape $r_i$, while each server $P_{i \in [n+1,N]}$ receives only an auxiliary input $z_i$ and random tape $r_i$. Each adversary $\adv_{i \in [m-1]}$ receives an index $i \in I $ that indicates the party it corrupts, while adversary  $\adv_m$ receives $C$ indicating the set of parties it corrupts.

For all $i \in H$, let $\OUT_i$ denote the output of honest party $P_i$ and, let $\OUT'_i$ denote the view of corrupted party $P_i$ for $i \in I \cup C$ during the execution of $\Pi$. The $i^{th}$ partial output of a real-world execution of $\Pi$ between parties $(P_1,\ldots, P_N)$ in the presence of adversaries $\adv = (\adv_1,\ldots, \adv_m)$ is defined as

$$\real^i_{\Pi,\adv, I, C, z_i,r_i}(k,x_i) \mydef \{\OUT_j \mid j \in H\} \cup \OUT'_i$$

\textbf{Ideal-world execution}. In the ideal-world execution, all the parties interact with a trusted party that evaluates a function $f$. Similar to the real-world execution, the ideal execution begins with each user $P_{i \in [n]}$ receiving its input $x_i$, an auxiliary input $z_i$, and random tape $r_i$. Each server $P_{i \in [n+1,N]}$ receives only an auxiliary input $z_i$ and random tape $r_i$. 

 Each user $P_{i \in [n]}$ sends $x'_i$ to the trusted party, where $x'_i$ is equal to $x_i$ if this user is semi-honest, and is an arbitrary value if he is malicious. If any semi-honest server sends an abort message ($\bot$), the trusted party returns $\bot$ to all users. The trusted party then returns $f(x'_1, \ldots, x'_n)$ to all the parties. 

For all $i \in H$, let $\OUT_i$ denote the output returned to the honest party $P_i$ by the trusted party, and  let $\OUT'_i$ denote some value output by corrupted party $P_i$ for $i \in I \cup C$. The $i^{th}$ partial output of a ideal-world execution of $\Pi$ between parties $(P_1,\ldots, P_N)$ in the presence of independent simulators $\Ss = (\Ss_1,\ldots, \Ss_m)$ is defined as

$$\ideal^i_{\Pi,\Ss, I, C, z_i, r_i}(k,x_i) \mydef \{\OUT_j \mid j \in H\} \cup \OUT'_i$$


\begin{definition} (Security)
\label{def:security}
	Suppose $f$ is a deterministic-time $n$-party functionality, and $\Pi$ is the protocol. Let $x_i$ be the parties' respective private inputs to the protocol. Let $I \in  [N]$ denote the set of corrupted and non-colluding parties and $C \in [N]$ denote the set of corrupted and colluding parties. We say that protocol $\Pi(I,C)$ securely computes deterministic functionality $f$ with abort in the presence of adversaries $\adv = (\adv_1,\ldots, \adv_m)$ if there exist probabilistic polynomial-time simulators $\Sim_{i \in m}$ for $m<n$ such that for all $\bar{x},\bar{z},\bar{r} \from \bool^\star$, and for all $i \in [m]$,	
	
	$$\{\real^i_{\Pi,\adv, I, C, \bar{z},\bar{r}}(k,\bar{x})  \widetilde{=}  \{\ideal^i_{\Pi,\Sim, I, C, \bar{z},\bar{r}}(k,\bar{x})\}$$
	Where  $\Ss = (\Ss_1,\ldots, \Ss_m)$ and $\Ss=\Sim_i(\adv_i)$
\end{definition}

\paragraph{Desirable Security/Privacy Properties.}
A desirable contract tracing system would make an honest user's actions perfectly indistinguishable from actions of all other honest users as well as servers. Thus, an ideal security system property would guarantee that executing the system in the real model is equivalent to executing this system in an ideal model with a trusted party as presented in the above definition \ref{def:security}.

Based on this definition of security, we consider the following attacks in the context of contact tracing.
\begin{itemize}
	\item Linkage attacks: A linkage attack attempts to match anonymized records with non-anonymized records in a different dataset~\cite{Dwork14}. For contract tracing there are two types of linkage attacks: by users and by the server.
	
	The adversarial server aims to link users and re-identify their contact history by observing tokens it receives. For example, if the server is able to deduce that Alice and Bob had come in contact, regardless of frequency or duration, that is a linkage attack, referred to in the Introduction as a social graph exposure.
	
	Even without connecting two users, if the server operator is able to track a single user over time, say by using Bluetooth beacons, that would also constitute a linkage attack.

	On the user side, most users are aware of who they are in contact with for at least some amount of time, thus linkage to tokens is not useful. Instead, we consider any other use of the anonymized information, such as identifying other users they do not already know, finding out about other users' contacts, the infection status of other users, or the source of their own exposure to the disease. Note that if a user was only near one individual during the infection period, and if she gets an alert of having been in contact with a confirmed case, then she knows who it was. This case cannot be avoided while providing the functionality of the application.
	
	\item False-positive claim: A malicious user may claim to have been diagnosed with the disease when in reality, they are not. This would spread false information and panic other users, and reduce trust in the system.
	
\end{itemize}

As we will demonstrate in the following sections, \dect provably provides all of the functions of contact tracing while protecting against the attacks above.

\section{Preliminaries}
This section introduces the notations and cryptographic primitives used in later sections.

\subsection{Notation}

In this work, the computational and statistical security parameters are denoted by $\kappa, \lambda$, respectively. For $n \in \mathbb{N}$, we write $[n]$ to denote the set of integers $\{1, \ldots, n\}$. We use `||' to denote string concatenation. We use party to refer to either a server or a user in the system.

\subsection{Cryptographic building blocks}
\label{subs:crypto_prelim}

\subsubsection{Decisional Diffie--Hellman}
\begin{definition}~\cite{Diffie:2006:NDC:2263321.2269104}
	 Let $\mathcal{G}(\kappa)$ be a group family parameterized by security parameter $\lambda$. For every probabilistic adversary $\adv$ that runs in polynomial time in $\lambda$, we define the advantage of $\adv$ to be:
	$$\abs{\Pr[\adv(g,g^a,g^b,g^{ab})=1]-\Pr[\adv(g,g^a,g^b,g^c)=1]}$$
	Where the probability is over a random choice \G from $\mathcal{G}(\lambda)$, random generator $g$ of \G, random $a, b, c \in [|G|]$ and the randomness of $\adv$. We say that the Decisional Diffie–Hellman assumption holds for \G if for every such $\adv$, there exists a negligible function $\epsilon$ such that the advantage of $\adv$ is bounded by $\epsilon(\lambda)$.
\end{definition}

\subsection{Pseudorandom Number Generator}
\label{subs:PRNG}
 
\begin{definition}~\cite{Koeune2011} A pseudorandom number generator (PRG) is a function that, once initialized with some random value (called the seed), outputs a sequence that appears random, in the sense that an observer who does not know the value of the seed cannot distinguish the output from that of a (true) random bit generator.
\end{definition}

\subsection{Discrete Log Zero-Knowledge Proof}
\label{subs:zk}
Discrete Log Zero-Knowledge Proof (DLZK) is a cryptographic  protocol, which allows Alice to convince Bob that she has $k$ for known $y=g^k$ in the cyclic group $\mathbb{G}=\langle g\rangle$ without revealing the value of $k$. One of the simplest and frequently used proofs of knowledge for discrete log is Schnorr protocol, which  incurs communication of $2$ group elements, and computation of $3$ modular exponentiations in a cyclic group.


\subsection{Garbled Circuits}
\label{subs:gc}
Garbled Circuit (GC) is currently the most common generic technique for two-party secure computation (2PC), which was first introduced by Yao\cite{FOCS:Yao86} and Goldreich et al.\cite{STOC:GolMicWig87}. The ideal functionality of GC is to take each party's inputs, $x$ and $y$ respectively, and compute some function $f$ on them. We denote this garbled circuit by $ z  \from \Fgc(x,y,f)$.  GC has seen dramatic improvements in recent years. Modern GC protocols~\cite{EC:ZahRosEva15, EC:WanMalKat17} evaluate two million AND gates per second on a 1Gbps LAN.  In our protocols, we use the ``subtraction" and ``less than"  circuits.

\remove{
\subsubsection{Homomorphic Encryption Scheme}
\label{subs:HE}
Homomorphic encryption is a form of encryption that allows to perform arbitrary computation on encrypted data, which consists of the following probabilistic polynomial-time algorithms:

\hgen($\lambda$): takes a security parameter $\lambda$ as input, and outputs a public-private key pair $(pk, sk)$

\henc($pk, m$): takes the public key $pk$ and a plaintext $m$ as input, and outputs a ciphertext  $ct=\henc(pk, m)$ as an encryption of $m$ under public-key $pk$.

\hdec($sk, ct$): takes the secret key $sk$ and a ciphertext $ct=\henc(pk, m)$ as input, and recovers the plaintext $m$.

\hsum{$(pk, \{ct_i\})$}: takes the public key $pk$ and a set of ciphertexts $\{ct_i=\henc(pk, m_i)\}$ as input, and outputs a ciphertext encrypting the sum of the underlying messages as $\henc{(pk, \sum\limits_{i}m_i)}$.

\hmul{$(pk, \{ct_i\})$}: takes the public key $pk$ and a set of ciphertexts $\{ct_i=\henc(pk, m_i)\}$ as input, and outputs a ciphertext encrypting the product of the underlying messages as $\henc{(pk, \prod\limits_{i}m_i)}$.

 In this work, we only focus on homomorphic encryption based on Ring Learning with Errors problem, which security derives from the difficulty of finding the shortest vector in a lattice. 
\paragraph{Multi-key FV Ring-LWE-Based homomorphic encryption.}
\todo{xxx}
}

\subsection{Private Information Retrieval}
\label{subs:pir}
Private Information Retrieval (PIR) allows a client to query information from one or multiple servers in a such way that the servers do not know which information the client requested. Recent PIR~\cite{EC:CacMicSta99,XPIRPrivateInformationRetrievalforEveryone,ESORICS:DonChe14,ICALP:GenRam05, SP:ACLS18} reduces communication cost to logarithmic in the database size.

In PIR, the server(s) hold a database $DB$ of $N$ strings, and the client wishes to read item $DB[i]$ without revealing $i$. 

\subsubsection{1-Server PIR}
In general, 1-server PIR construction~\cite{10.1007/978-3-319-11203-9_22,SP:ACLS18,  EPRINT:ALPRSSY19} consists of the following algorithms: 
   
 \begin{itemize}
 	
 	\item $\pirge(\kappa) \to (pk,sk)$: takes a security parameter and generates an additively homomorphic public and secret key pair $(pk,sk)$. 
 	
 	\item $\pirqe(pk, i) \to k$: a randomized algorithm that takes index $i \in [N]$ and public key $pk$ as input and outputs a (short and unexpanded) key $k$ of size $O(\log(N))$ bits or $O(d \ceil{\frac{\sqrt[d]{N}}{p}})$ bits, where $d$ is typically 2 or 3, and  $p$ is typically 2048 or 4096. 
 	
 	\item $\pirexpand(pk, k) \to K$: takes a short key $k$ and public key $pk$ as input and outputs a long {\em expanded key} $K \in \{0,1\}^N$.%
 	
 	\item $\pirans(pk, K, DB) \to d$: takes an expanded key $K$, public key $pk$, and a database $DB$ as input, returns an answer $d$ encrypted under $pk$.
 	
 	\item $\pirextract(sk, d) \to DB[i]$: takes a secret key $sk$ and answer $d$ as input, returns $DB[i]$.
 \end{itemize}

With these algorithms defined, PIR generally proceeds as follows.

\begin{enumerate}
    \item The client generates a public and secret key pair with \pirge. This is generally done once at setup and then reused.
    
    \item 
    The client uses \pirqe\ to generate a query key $k$ for the desired item, and sends both $k$ and $pk$ to the server.
    
    \item The server uses \pirexpand\ to expand $k$ to the much larger $K$, and then uses \pirans\ to generate the answer $d$, which is transmitted to the client.
    
    \item The client then reconstructs $DB[i]$ using \pirextract.

\end{enumerate}

PIR is generally implemented such that if $(k) \gets \pirqe(i)$, then $K \gets \pirexpand(k)$ is the encryption of a binary string of zero everywhere except for a 1 in the $i$th bit under public key $pk$. \pirans\ then consists of iterating over all of the items in the database and computing $d \overset{\textrm{def}}= \textstyle  \bigoplus_{j=1}^N K[j] \cdot DB[j]$. Thus, $d$ is the encryption of $DB[i]$ under the public key $pk$.

Most single-server PIR constructions~\cite{10.1007,10.1007/978-3-319-11203-9_22,SP:ACLS18,  EPRINT:ALPRSSY19} have communication cost of either $O(\log(N))$ bits or $O(d \ceil{\frac{\sqrt[d]{N}}{p}})$ bits, where $d$ is typically 2 or 3, and  $p$ is typically 2048 or 4096. Depending on the implementation chosen, the latter may actually be faster for the application due to the size of $d$, $p$, and constant values. Single-server PIR requires computation of $O(N)$ additive homomorphic operations on the server. There is also a tradeoff between communication and computation costs, as discussed in~\cite{EPRINT:ALPRSSY19}, which must be evaluated for optimal performance in the application.  We provide more analysis of performance in \sectionref{subs:perf}.

\subsubsection{2-Server PIR}
The $O(N)$ homomorphic operations required for single server PIR can be very expensive for large databases and serving large userbases. In order to reduce the computational overhead on the server's side, some PIR schemes use multiple servers with the assumption that not all of them collude \cite{EC:BoyGilIsh15,CCS:BoyGilIsh16}.

So-called 2-server PIR replaces homomorphic encryption with symmetric encryption (typically AES) and bit operations, making the following changes to the single server scheme:

\begin{enumerate}
    \item \pirqe\ uses a PRF (such as AES) to produce two query keys, $k_1, k_2$, each of size $O(\log(N))$, which the client sends to server 1 and server 2 respectively.
    
    \item Each server then expands their key by $K_i \gets \pirexpand(k_i)$. An important property of the keys is that $K = K_1 \oplus K_2$ is zero everywhere except for position $i$ which is 1. Since each server only has one of the two $K_i$ values, and assuming the two servers do not collude, they cannot determine the value of $K$ or $i$.
    
    \item Both servers then locally compute the inner product $d_i \overset{\textrm{def}}= K_i \cdot DB = \textstyle\bigoplus_{j=1}^N K_i[j] \cdot DB[j]$, and then send the result to the client.
    
    \item The client can then reconstruct $d = d_1 \oplus d_2 = (K_1 \cdot DB) \oplus (K_2 \cdot DB) = (K_1 \oplus K_2) \cdot DB = DB[i]$
    
\end{enumerate}

In 2-server PIR~\cite{EC:BoyGilIsh15,CCS:BoyGilIsh16}, the communication cost is $O(\log(N))$ bits and the computation requires $O(N)$ symmetric key operations which is much faster than additive homomorphic operations.

\subsubsection{Keyword PIR}

Chor, et al.~\cite{EPRINT:ChoGilNao98} define a variant of PIR called keyword PIR, in which the client has an item $x$, the server has a set $S$, and the client learns whether $x \in S$. This variant of PIR has been used for the password checkup problem~\cite{EPRINT:ALPRSSY19}, where a client aims to check whether their password is contained in breached data, without revealing the password itself. One implementation of keyword PIR is based on PIR with Cuckoo hashing~\cite{10.1007}, which requires approximately three times the costs of regular PIR. Another solution~\cite{EPRINT:ALPRSSY19} relies on bucketing which we describe more detail in Section \ref{subs:impl_optz}. In this paper, we are interested in Keyword PIR based on both 1-server PIR~\cite{SP:ACLS18,  EPRINT:ALPRSSY19} and 2-server PIR~\cite{EC:BoyGilIsh15,CCS:BoyGilIsh16}.

\subsection{Private Set Intersection Cardinality}	
Private set intersection cardinality (\psica) is a two-party protocol that allows one party to learn the intersection size of their private sets without revealing any additional information~\cite{NDSS:HuaEvaKat12, USENIX:PSSZ15,EC:PSWW18, EC:PSTY19,EPRINT:IKNPRSSSY19}. The \psica functionality is presented in \figureref{fig:psica-func}.

\begin{figure*}[t!]
	\centering
	\framebox{
		\begin{minipage}{0.95\linewidth}
			{\sc Parameters:}
			Two parties: server and client; and upper bound on the input set size.
			
			
			{\sc Functionality:}
			\begin{itemize}[noitemsep,topsep=0pt]
				\item Wait for input set $X$  from the server
				\item Wait for input set $Y$ from the client
				\item Give server nothing		
				\item Give client $|X \cap Y|$		
			\end{itemize}
		\end{minipage}
		}
	\caption{The \psica Functionality.}
	\label{fig:psica-func}	
\end{figure*}

\section{Our \dect System}
\label{sect:system}
We now present the \dect system in detail, the construction of which closely
follows the high-level overview presented in \sectionref{subs:system_overview}. Recall that \dect aims to alert any users who have, within the infection window (e.g. 14 days for COVID-19), come into contact with another user who has been diagnosed positive with an infectious disease.

\subsection{System Phases}

\dect's design combines several different cryptographic primitives. In this section we will describe how \dect uses these primitives to obtain the goals described in \sectionref{sec:goals}. Please refer to \sectionref{subs:crypto_prelim} and \sectionref{sect:psi_prot} for more details on the cryptographic gadgets used here. The \dect system consists of four phases as follows.

\subsubsection{Agreement and Setup Phase}  

The first phase requires all parties (including users, the healthcare provider, and the \dect server) to agree to perform the objective function (e.g. contact tracing) over their joint data, and security parameters for MPC.  The parties should also agree to release the computed result to each user. This agreement might happen before the user installs \dect on their phone.

The \dect server takes a security parameter $\lambda$ as input, and outputs a public-private key pair $(pk, sk)$, and shares the public key with every user. Each user/client $u_i$ generates a random PRG seed $s_i$ which it uses to generate contact tokens in the next phase. 
As long as the server's configuration does not change, this phase does not need to be re-run. Whenever a new user registers with \dect, they only needs to generate their own PRG seed, and the server shares the public key with the new user. 

\begin{figure*}[t]
	\centering
	\includegraphics[scale=0.51]{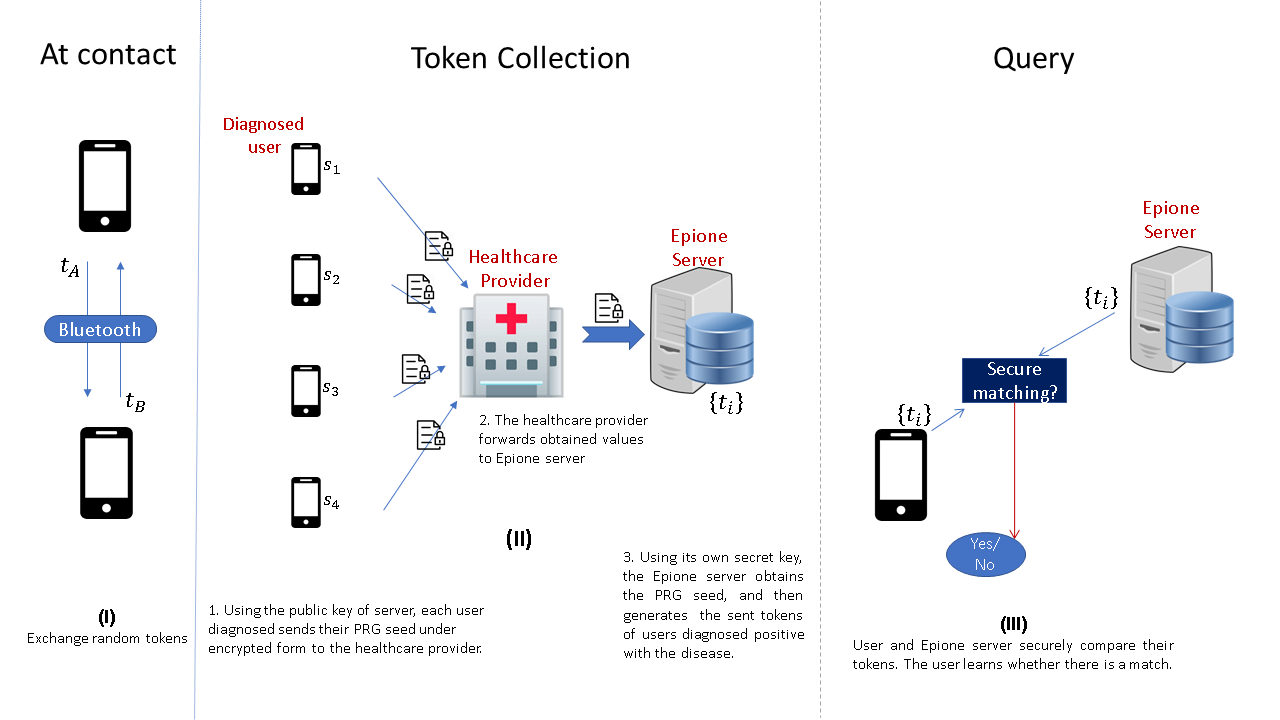}
	\caption{\dect System Design without Agreement and Setup Phase. (I) Tokens are exchanged when two users are in close proximity. (II) When a user is diagnosed with the disease, the user encrypts their PRG seed using the public key of the \dect server, and gives the encrypted value to the healthcare provider, who then transmits it to the \dect server. Using its private key, the \dect server decrypts the received ciphertexts and obtains the PRG seeds of diagnosed users. The \dect server generates the sent tokens of users diagnosed positive using the PRG. (III) Each user invokes a secure matching algorithm with the \dect server, where the user's input is their received tokens and the server's input is the database of tokens from users diagnosed with the disease. The user learns only whether (or how many) tokens there are in common between the two sets, while the \dect server learns nothing.
	}
	\label{fig:system}
\end{figure*} 

\subsubsection{Token Generation}

Similar to most recent contract tracing systems~\cite{cho2020contact,Covid-watch,CEN,chan2020pact}, we use Bluetooth to exchange contact tokens whenever two users are in close proximity. The $\ctgenerate(s_u, d_i, j) \to t_{u,i,j}$ function is used to generate tokens of $\kappa$ bits each to be sent by user $u$ on day $i$ and timeslot $j$. The precise details of the token generation are left as an implementation detail, so long as the following criteria are met:

\begin{itemize}
    \item Tokens are indistinguishable from random by anyone not in possession of the user's seed $s_u$. In other words, the \ctgenerate\ function acts as a PRG as defined in \sectionref{subs:PRNG}.
    
    \item Tokens can be deterministically generated for the given day $d_i$, and time slot $j$ using a secret seed, $s_u$, such that when a user gives their seed to the \dect server, the server is able to regenerate the tokens sent by the user.
    
    \item All users and the \dect server agree on the method used to generate tokens, the time intervals, and day numbering.
\end{itemize}

\subsubsection{Contact}

As illustrated in \figureref{fig:system} part I, when two users, say Alice and Bob, enter within close proximity, \dect detects this condition with a short range network such as Bluetooth, and then uses that network to exchange tokens using the function \ctexchange. Alice generates token $t_a \gets \ctgenerate(s_a, d_i, j)$, where $s_a$ is Alice's private seed, $d_i$ is the current day, and $j$ is the current time slot. Similarly Bob generates token $t_b \gets \ctgenerate(s_b, d_i, j)$. Alice sends $t_a$ to Bob, and Bob send $t_b$ to Alice.

Alice then adds the token received from Bob, $t_b$, to her set of received tokens, $\sT_{R,A}$, and Bob adds $t_a$ to $\sT_{R,B}$. We use $\sT_{S,A}$ to represent the set of tokens Alice has sent to other users (which includes $t_a$), though Alice does not actually store such a list since it can be regenerated at any time from her private seed. Alice and Bob discard received tokens that are older than the infection window (e.g. 14 days for COVID-19).

\subsubsection{Positive Diagnosis and Token Collection}

When a user ($u_i$ in general) is diagnosed with the disease, the user encrypts their PRG seed using the public key of the \dect server and gives that to the healthcare provider (provided the user consents to this, of course). The healthcare provider gathers the seeds from several users diagnosed positive, shuffles them, and transmits the set of seeds over a secure channel to the \dect server. Using its private key, the \dect server decrypts the received values to obtain the secret PRG seeds. The \dect server can then generate all of the tokens for the infection window sent by users diagnosed positive with the disease, $\widehat{\sT}_S$. The token collection process is shown in part II of \figureref{fig:system}.

Two servers are used at this phase to prevent any one server from knowing both the diagnosis status of a user and their sent tokens. This is useful in the case that the \dect server is operated by an untrusted party, such as a commercial provider, that should not have access to sensitive information such as a user's diagnosis. If such protection is not needed, for example if the \dect server is operated by a health authority that already has access to the infection status of users and can be trusted not to try to discern a user's diagnosis status from the token collection process, then both services can be provided by the same server.

Alternatively, the healthcare provider could provide a token to the user that the user then provides the \dect server when they upload their tokens to prove that they have a legitimate positive diagnosis. This would allow the \dect server to verify that the user's claim is legitimate, but does not protect the user from the server linking them to a positive diagnosis.

\subsubsection{Query}
\label{sub:pect_psi}

Recall from the contact phase that each user $u_i$ keeps a list of tokens received from other users they have been in contact with within the infection window, $\sT_{R,u_i}$. The query phase aims to securely compare the user's received contact tokens $\sT_{R,u_i}$ with the \dect server's set of tokens sent by users diagnosed positive with the disease, $\widehat{\sT}_S$. If there are any tokens in common, then user $u_i$ has come into contact with an individual diagnosed positive within the infection window, and should be notified that they are at risk of having contracted the disease. This process is illustrated in part III of \figureref{fig:system}.

The comparison of tokens is done by calling the \ctquery\ function, which we implement using \psica. We describe \psica\ in detail in \sectionref{sect:psi_prot}. Note that revealing the intersection size is acceptable in the contact tracing application we consider, however, it is possible hide the intersection size as we discuss in \sectionref{subs:tpsi}.

\remove{	
\begin{figure*}[ht!]
	\centering
		\noindent\fcolorbox{black}{gray!20}{%
		\begin{minipage}{0.95\linewidth}
			{\sc Parameters:}
			Two parties: server and user; and upper bound on the input size of each party.
			
			
			{\sc Functionality:}
			\begin{itemize}[noitemsep,topsep=0pt]
				\item Wait for input set $\widehat{\sT}_S$  from the server
				\item Wait for input set $\sT_{R,u_i}$ from the user
				\item Give nothing to the server 
				\item Give the user a bit indicated whether $|T_i \cap \sT|>0$		
			\end{itemize}
		\end{minipage}
	}
	\caption{Our \tpsi Gadget.}
\label{fig:tpsi-func}	
\end{figure*}
}

\subsection{Security Discussion}
\label{subs:security_discussion}

In this section we consider the security of \dect, starting from a general theorem and then considering specific attacks and protections as defined in \sectionref{subs:probl}.

\subsubsection{Security Theorem}

The security of \dect follows in a straightforward way from the security of its building blocks (e.g. \psica) and the PRG scheme. Thus, we omit the proof of the following theorem.

\begin{theorem}
	\label{thm:dect}
	The \dect construction securely implements the contact tracing functionality defined in \sectionref{subs:probl} in the semi-honest setting, given the \psica primitive described in Figure \ref{fig:psica-func} and a secure pseudo-random number generator (PRG) as defined in \sectionref{subs:PRNG}.
\end{theorem}

\subsubsection{Defense against linkage attacks}

As defined in \sectionref{subs:secure_def}, a linkage attack is any attempt to link an anonymized record with any identifying information. \dect successfully defends against all important linkage attacks.

\paragraph{Linkage attacks by server.} The \dect server has only the tokens sent by users who have been diagnosed with the disease. It cannot identify which users have been diagnosed with the disease without colluding with the healthcare provider (the latter is assumed to already know such information, and is responsible for verifying that the user's diagnosis is legitimate). Similarly, the healthcare provider does not have access to even randomized contact tokens without colluding with the \dect server.

Even if the \dect server is able to link a user to specific tokens or seeds in its database, the server does not gain any further information. This limits the amount of exposure in the event of collusion or a breach of the database.

By using \psica, \dect prevents any kind of social graph exposure. The \dect server has only a set of randomized sent contact tokens, and does not know which users have received those tokens. With \psica, users query for the number of tokens they have in common between the \dect server set and their own received tokens set, without the server gaining any knowledge of their received tokens.

\paragraph{Linkage attacks by users.} Because users only learn the number of contact tokens they have received from users since diagnosed with the disease and not the tokens themselves, semi-honest users cannot link their exposure to the disease to any particular user.

\paragraph{User tracking and identification.} Because tokens appear random, users cannot be tracked using Bluetooth beacons, whether by other users or the \dect server. It's also impossible to identify a user based solely on a token received from that user without extra information.

\subsubsection{Malicious User Queries}

If a malicious user, Mallory, can deviate from the protocol by submitting arbitrary queries to the \dect server, it is possible for her to craft queries in such a way as to perform a search, and find which token(s) in her set are also in the server's set. If Mallory also records the time, place, and who she was with for the tokens she has received, she can later use this information to glean which user(s) have been diagnosed with the disease.

There are several ways to mitigate the threat of arbitrary queries. First, 
we could require that users submit a cryptographic hash (e.g. by computing a Merkle root\footnote{The following details ensure that the committed hash value is randomized and defeats any dictionary attacks by the server: We first permute the local tokens randomly, add a dummy random value in the list, and then compute a Merkle tree. The Merkle root is committed as the cryptographic hash of the list. The adversary does not know the dummy value or the permuted order and hence is unable to forge a Merkle proof.}) of their local token
list periodically to the \dect server, say once every day.
When Mallory queries the server, her query includes the cryptographic hash of the set used in the query. The hash ensures that the entire set of tokens obtained by Mallory at a particular point in time is used in the query, and not some chosen subset. Because the query items are encrypted\footnote{Either under DDH for token transformations, or homomorphic encryption for 1-server PIR, or AES for 2 server PIR, depending on at which step the check is done}, the \dect server cannot directly verify that the hash is correct. Instead, Mallory must also provide a zero knowledge proof that the query items correspond to the hash submitted, such as the SNARK method proposed in~\cite{USENIX:BCTV14}. Using this, the \dect server verifies that the query matches the hash.

Periodic commitment to cryptographic hashes of a user's token set mitigates the threat of false claims as well. Users cannot retroactively add tokens to their local lists without being detected by the \dect server.

Note that the contact tracing functionality itself reveals whether there is a match within Mallory's full set of tokens. This is not a linkage attack, but a direct implication of the desired functionality of the application. Mallory, or any benign user, may have had contact with only one other user during the infection window, in which case they can deduce the infection status of the other user when they query the \dect server for intersection cardinality. This is fundamentally unavoidable as it is part of the benign functionality of the application.

A second complementary mitigation is rate limiting users to a few queries per day, and requiring a minimum number of tokens per query. By itself that would slow Mallory down, but not prevent the attack completely. With enough queries, Mallory will eventually deduce with high probability the source of her exposure.

Lastly, using Digital Rights Management (DRM) protocols such as Android SafetyNet and Apple DeviceCheck will make it much harder for Mallory to submit such queries and is highly recommended. When combined with rate limits and a minimum token set size, these protocols make crafting queries to find the exposure source impractical, though again not impossible. 
\section{Cryptographic Gadgets}
\label{sect:psi_prot}

This section provides more detail on the cryptographic tools we use to implement \dect, with a specific emphasis on our \psica design and PIR, as well as extensions to those tools.

\subsection{PSI cardinality (PSI-CA) for asymmetric set sizes}
\label{subs:psica}

In this section, we present our \psica construction, the functionality of which is described in \figureref{fig:psica-func} and used as a core component of \dect.

\subsubsection{Our technique}

We start with a private set intersection (PSI) in the semi-honest setting, where two parties want to learn the intersection of their private set, and nothing else. The earliest protocols for PSI were based on the Diffie-Hellman (DH) assumption in cyclic groups. Currently, DH-based PSI protocols~\cite{Huberman99} are still preferable in many real-world applications due to their low communication cost.

\begin{figure*}[h]
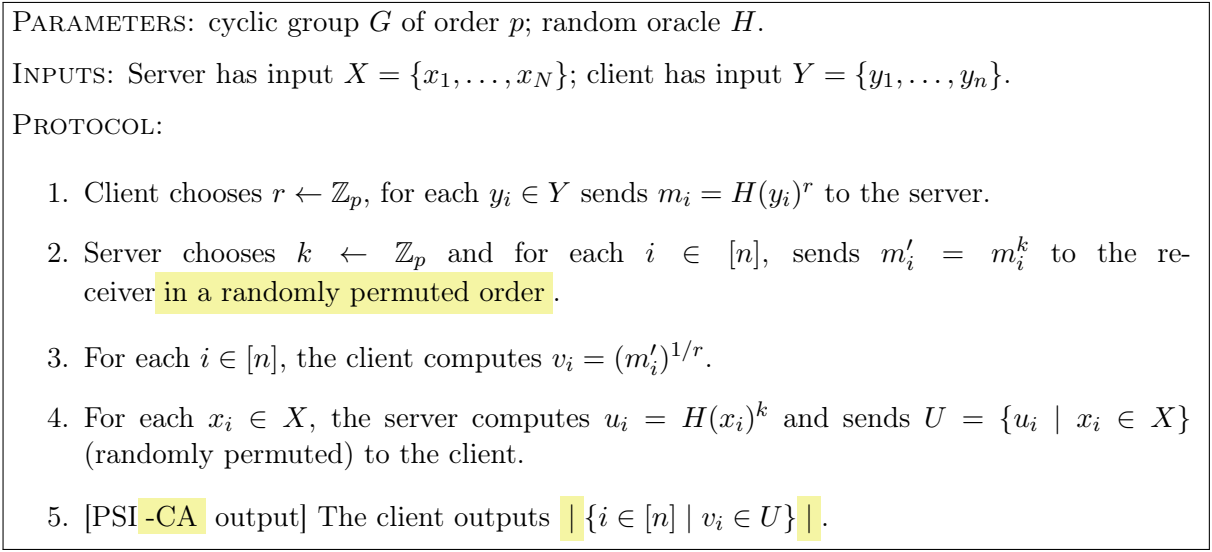

	\centering
	\framebox{\begin{minipage}{0.95\linewidth}
			{\sc Parameters:}
			cyclic group $\G$ of order $p$; 
			random oracle $H$.
			
			\medskip
			
			{\sc Inputs:} Server has input $X = \{x_1, \ldots, x_{N}\}$; client has input $Y = \{ y_1, \ldots, y_{n} \}$.
			
			\medskip
			{\sc Protocol:} 
			
			\begin{enumerate}
				\item 
				Client chooses $r \gets \Z_p$, for each $y_i \in Y$ sends $m_i = H(y_i)^r$ to the server.
				
				\item
				Server chooses $k \gets \Z_p$ and for each $i\in [n]$, sends $m'_i = m_i^k$ to the receiver\basehighlight{in a randomly permuted order}. 
				
				\item For each $i \in [n]$, the client computes $v_i = (m'_i)^{1/r}$.
				
				\item
				For each $x_i \in X$, the server computes $u_i = H(x_i)^k$ and sends $U=\{ u_i \mid x_i \in X \}$ (randomly permuted) to the client.
				
				\item{} [PSI\basehighlight{-CA} output] The client outputs $\mathhighlight{|}\{ i \in [n] \mid v_i \in U \}\mathhighlight{|}$.
			\end{enumerate}
		\end{minipage}
	}
	\caption{DH-based PSI protocol and  extension to PSI-CA \basehighlight{with changes highlighted}.}
	\label{fig:dh-psi}
\end{figure*}

\paragraph{DH-based PSI}
 Assume that the server has input $X = \{x_1, \ldots, x_{N}\}$ and client has input $Y = \{ y_1, \ldots, y_{n} \}$.  Given a random oracle $H : \{0,1\}^* \to \G$, and a cyclic group $\G$  in which the DDH assumption holds, the basic DH-based PSI protocol is shown in \figureref{fig:dh-psi}.
 
Intuitively, the client sends $\{ H(y_i)^r \}_{y_i \in Y}$ for some random, secret exponent $r$. The server raises each of these values to the $k$ power, and the client can then raise these results to the $1/r$ power to obtain $\{ H(y_i)^k \}_{y_i \in Y}$ as desired\footnote{Alternatively, the client can raise each $H(x_i)^k$ to the $r$ power and compare to the $H(y_i)^{kr}$ values. However, the variant where the client raises $H(y_i)^{kr}$ to the $1/r$ power is compatible with further optimizations.}.

\paragraph{From DH-based PSI to PSI cardinality (PSI-CA).}

If the client uses the same $r$ for every item, it is possible to extend the basic PSI algorithm to compute functions such as intersection set size (cardinality) without revealing the intersection items by having the server shuffle the items. This observation was suggested by \cite{Huberman99} and recently was incorporated into private intersection sum~\cite{EPRINT:IKNPRSSSY19}, which allows two parties to compute the sum of payloads associated with the intersection set of two private datasets, without revealing any additional information. Clearly, PSI-CA is a special case of private intersection sum, where the payload is constant and equal to 1. 

\figureref{fig:dh-psi} also shows the extension to PSI-CA with the highlighted changes. The key idea to transform PSI into PSI-CA is that instead of sending $m'_i$ in step 2 of \figureref{fig:dh-psi} in order, the server shuffles the set in a randomly permuted order.  Shuffling means the client can count how many items are in the intersection (PSI-CA) by checking whether  $v_i \in U$, but learns nothing about which specific item was in common (e.g. which $v_i$ corresponds to the item $y_j$). Thus, the intersection set is not revealed.

\paragraph{From PSI-CA to PSI-CA for asymmetric sets.}

In many applications, including contact tracing, the two parties (client and server) have sets of extremely different sizes. A typical client has less than 500 new tokens per day, while the server may
have millions of tokens in its input set. In PSI, most work is optimized for the case where two parties have sets of similar size, and as such their communication and computation costs scale with the size of the larger set. For contact tracing, it is crucial that the client's effort (especially communication cost) be sub-linear in the server's set size. More practically, we aim for communication of at most a few megabytes in a setting where the client is a mobile device. 

We observe that the last two steps of \figureref{fig:dh-psi} are similar to the function performed by keyword PIR, which is communication-efficient in the conventional client-server setting. Keyword PIR allows clients to check whether their item is contained in a set held by a server, without revealing the actual item to the server. Therefore, step 4 and 5 of \figureref{fig:dh-psi} can be replaced by keyword PIR. Concretely, after step 3, the client has an input set $V=\{v_1, \ldots, v_n\}$ and the server has input set $U=\{u_1, \ldots, u_N\}$. The client sends a multi-query keyword PIR request with all of the elements in $V$ to be queried against $U$ on the server. From the PIR response, the client can count the number of $v_i \in U$ to find the set size, without revealing to the server the actual values in $V$ and without the client learning any more information about $U$.

\subsubsection{Protocol}
Our semi-honest PSI-CA protocol is presented in \figureref{fig:unbalance_psica}, following closely the description in the previous subsection. The client runs keyword PIR searches for each $v_{i\in [n]}$ in a set $U$ held by the server. For communication and computation efficiency, the values of both $u_i$ and $v_i$ can be truncated, and the protocol will still be correct as long as there are no spurious collisions. We can limit the probability of such a collision to $2^{-\lambda}$ by truncating to length $\lambda + log(N)$ bits. In \figureref{fig:unbalance_psica}, we use a truncation function $\tau(z)$ which takes $z$ as input and returns the most significant $\lambda + log(N)$ bits of $z$.

\begin{figure*}[h]
	\centering
	\framebox{\begin{minipage}{0.95\linewidth}
			{\sc Parameters:}
			cyclic group $\G$ of order $p$; random oracle $H$, Multi-query Keyword PIR primitive (\sectionref{subs:pir}), a truncate function $\tau(z)$ takes $z$ as input and returns first  $\lambda + log(N)$ bits of $z$.
			
			\medskip
			
			{\sc Inputs:} Server \basehighlight{1} has input $X = \{x_1, \ldots, x_{N}\}$; client has input $Y = \{ y_1, \ldots, y_{n} \}$; \basehighlight{Server 2 has no input}
			
			\medskip
			{\sc Protocol:} 
			
			\begin{enumerate}
			
				\item Server \basehighlight{1} chooses $k \gets \Z_p$, and computes dataset  $U=\{ \tau\big(H(x_i)^k\big) \mid i \in [N] \}$. \basehighlight{Server 1 sends $U$ to Server 2}
				
				\item Client chooses $r \gets \Z_p$, and sends $m_i = H(y_i)^r$ for each $y_i \in Y$ to the server \basehighlight{1}.
				
				\item
				Server \basehighlight{1} chooses a random permutation $\pi: [n] \to [n]$. For each $i\in [n]$, sends $m'_i = (m_{\pi(i)})^k$ to the client.
				
				\item For each $i \in [n]$, the client computes $v_i = \tau\big((m'_i)^{1/r}\big)$.
				
					\item Parties invoke Multi-query Keyword-PIR \basehighlight{with 2 servers}:
				\begin{itemize}
					\item Server \basehighlight{1} acts as Keyword-PIR server \basehighlight{1} with dataset $U$
					\item\basehighlight{Server 2 acts as Keyword-PIR server 2 with dataset $U$}
						
					\item Client acts as Keyword-PIR client with $V=\{ v_i \mid i \in [n] \}$
					\item Client learns whether $v_i \in U, \forall i \in [n]$
				\end{itemize}
				
			\item Client outputs $|V \cap U|$
			
			\end{enumerate}
		\end{minipage}
	}
	\caption{Our semi-honest PSI-CA protocol for asymmetric sets, and extension to 2-server PIR based PSI-CA \basehighlight{with changes highlighted}}
	\label{fig:unbalance_psica}
\end{figure*}

\begin{theorem}
	\label{thm:psica}
	The PSI-CA construction of \figureref{fig:unbalance_psica} securely implements the \psica functionality defined in \figureref{fig:psica-func} in semi-honest setting, given the Multi-query Keyword PIR described in \sectionref{subs:pir}.
\end{theorem}

\begin{proof_sketch}

	The security of the PSI-CA protocol follows from the fact that $\{ H(z_1)^k, \ldots, H(z_n)^k \}$ are indistinguishable from random, for {\em distinct} $z_i$, if $H$ is a random oracle and the DDH assumption holds in $\G$.
	To see why, consider a simulator that receives a DDH challenge $g^k, g^{a_1}, \ldots, g^{a_n}, g^{c_1}, \ldots, g^{c_n}$ where each $c_i$ is either random or $c_i = a_i k$.
	The simulator programs the random oracle so that $H(z_i) = g^{a_i}$ and then simulates the outputs as $\{ g^{c_1}, \ldots, g^{c_n} \}$.
	It is easy to see that these messages are distributed as specified by $F$ if the $c_i$'s are distributed as $a_i k$, but are distributed uniformly otherwise, with the difference being indistinguishable by the DDH assumption.
	
	We consider two cases, corresponding to each party being corrupt.
	\begin{itemize}
		\item A corrupt server first sees $\{ H(y_i)^r\}_{y_i \in Y}$. By our observation above, these values are pseudorandom. A corrupt server also sees PIR transcripts. Because pseudorandomness guarantees of PIR,  the client's message to the server can be simulated as a random message. 
		
		\item A corrupt client sees $\{ H(y_{\pi(i)})^{rk} = (H(y_{\pi(i)})^k)^r \}_{y_{\pi(i)} \in Y}$  and PIR response (as the extra PSI message), along with its private randomness $r$ and any random oracle queries/responses that it made.
		Consider modifying this view, replacing each  $H(z)^k$ term with an independently random group element for each $z \in X \cup Y$ (each distinct, by definition).
		This change will be indistinguishable, by the reasoning above.
		Now it is not necessary to know the identities of $x_i \in X \setminus Y$ as well as $x_i \in X \cup Y$, as their corresponding $H(x_i)^k$ values have been replaced with random group elements that are independent of everything else, and the secret permutation $\pi$ hides the common items.
		In other words, this is a distribution that can be generated by the simulator, with knowledge of only $Y$ and $|X \cap Y|$.
	\end{itemize}
	
\end{proof_sketch}

\subsubsection{PSI-CA Cost} 

In our PSI-CA protocol, communication cost is $O(n\log(N))$ while the client's computation is $O(n)$ and the server's computation is $O(nN)$. More specifically,

\begin{itemize}
	\item The server and client must communicate (1) $O(n)$ group elements, (2) $O(n)$ homomorphically encrypted selection vectors for Keyword-PIR. If Keyword-PIR uses $O(\log(N))$ bits for each vector\footnote{there is a tradeoff between communication and computation complexity in PIR/Keyword-PIR as discussed in \cite{EPRINT:ALPRSSY19}. Traditional PIR is $O(\log(N))$ for query vectors, but some schemes trade slightly higher communication complexity for reduced computational complexity.}, the total communication cost is $O(n\log(N))$ bits.  We provide more analysis of performance in \sectionref{subs:perf}.
	
	\item The client's computation cost consists of: (1) $O(n)$ group elements, (2) $O(n)$ homomorphic encryptions and decryptions for encoding the Keyword-PIR queries and decoding the results.
	
	\item The server's computation cost consists of: (1) $O(n)$ group elements, (2) $O(nN)$ additive homomorphic encryption operations for finding the answer to the Keyword-PIR query.

\end{itemize}

The two-server PIR model can be used to speed up the server side computation by avoiding homomorphic encryption operations.

\subsection{PSI-CA with 2-server PIR}
Recall that the client and server invoke Keyword PIR in step 5 of~\figureref{fig:unbalance_psica}. To speed up the computational overhead on the server side, we introduce a second, independently operated server. The primary server sends the dataset $U$ to the second server after it has been computed. By DDH, the second server learns nothing about the item $x_i$ from $u_i$.

The client sends PIR queries with keyword $v_i$ to both servers, and learns whether $v_i \in U$ and nothing else. Neither PIR server learn anything about the client's query as long as the two servers do not collude. 

With 2-server PIR, the computation cost of PIR contains only symmetric-key operations, using approximately $2N$ PRF calls, and the communication cost of PIR is $O(log(N))$ bits. The highlights in \figureref{fig:unbalance_psica} shows the changes in \psica to go from single-server to 2-server PIR.

\subsection{PSI-CA Extensions}

Here we discuss two extensions to our \psica protocol to provide additional protections.

\subsubsection{Potential approach from PSI-CA into  threshold PSI-CA (\tpsi)}
\label{subs:tpsi}

Revealing the intersection size is acceptable in the contact tracing application we consider. However, it is possible that in other settings, knowing the size of the intersection is undesirable leakage.

Threshold PSI-CA is an extension of PSI where parties learn the intersection size (or even the intersection items) if it is greater than a given threshold. In our \tpsi definition, the client learns whether two input sets have any common items and nothing else (e.g. $t=0$). 

A simple solution is to pad both input sets with dummy elements. The two parties decide on a PRG seed $s$ which is used to generate common dummy elements. Each party randomly chooses a number of dummy elements, $n'$ and $N'$ for client and server respectively such that $n' > N'$. This can be done by performing the following steps:

\begin{enumerate}
    \item Parties randomly choose $n'$ and $N'$
    \item Parties invoke a garbled circuit to check whether $n' > N'$
    \item Repeat this process until the ``if'' condition is true
\end{enumerate}

The client and server then use the agreed PRG and seed $s$ to generate $n'$ and $N'$ fake items, respectively, and add them to their input sets.  The resulting intersection set size over the original and common dummy elements will be $\sigma=|X \cap Y|+N'$. The client learns $\sigma$ at the end of \psica, but does not know $N'$, and thus has limited knowledge of $|X \cap Y|$.

The parties then invoke a garbled circuit to securely remove the term $N'$ in $\sigma$ and check whether $|X \cap Y| > t$. The circuit takes as input $N'$ from the server and $\sigma$ from the client, computes $f=\big(\sigma- N'\big) >t?$, and returns the result to the client.

There remains an important concern in this approach: how to choose $n'$ and $N'$, so that $\sigma$ informationally hides the actual intersection size. Since the range of intersection size is from $0$ to $n$, the bound of information leakage is $O(\log(n))$ bits. Therefore, it is sufficient to choose $n'$ and $N'$ to be  $O(2^{\log(n)})$, which is essentially $O(n)$. However, it is not clear what the coefficient value behind the big O needs to be to prevent leakage of the actual intersection size. For example, the client can infer the lower bound of the intersection size by knowing that  $|X \cap Y| = \sigma - N' > \sigma - n'$. Further analysis needs to be done to prove if this method provides sufficient bits to prevent information leakage. 


\subsubsection{Potential approach for extending to malicious client}
\label{subs:malicious_user}

In the context of contact tracing, a malicious client seeks to obtain information about the server's database (set $X$ in this case). Thus, they attempt to compute $m_i = H(y_i)^r$ incorrectly, since in step 3 of \figureref{fig:unbalance_psica} the server returns $m'_i=(m_{\pi(i)})^k$ and it may be possible to learn part or all of the value of $k$. With that, the client can determine which values of $v_i$ map to which values of $u_j$, and thus learn which items from its set exist on the server.

One solution to prevent this is to augment the protocol with a zero-knowledge proof~\cite{SCN:JarLiu10,AC:DeCKimTsu10} that the $m_i$'s were computed correctly.
This adds the following step to the protocol:
\begin{itemize}
	\item[(2a)] Client performs a zero-knowledge proof of knowledge of $r$ such that $\forall i \in [n]: m_i = (y_i)^r$. The server aborts if the proof does not verify.
\end{itemize}

This modification is enough to guarantee security against a malicious client for the DH-based PSI protocol. 

Achieving a \psica or \tpsi protocol resilient to a malicious client requires more work because the construction of the client's message involves more building blocks, namely keyword PIR and garbled circuit. The solution is to use versions of these building blocks resilient to such one-side malicious attacks. Moreover, to verify that the client uses the correct values for $v_i$ for step 5 of \figureref{fig:unbalance_psica} in the keyword PIR query, we employ a consistent check such as  MACs from the SPDZ protocol~\cite{C:DPSZ12}. We will explore more detail in this direction.

\section{Implementation Choices and Performance Estimates}
\label{subs:perf}

In this section we consider some important implementation decisions and estimate the performance of \dect in order to show that the system is feasible in practice. While some performance optimizations are assumed in order to make the system realistic, we aim to be conservative with our estimates in order to give a loose upper bound to what can be expected for a practical implementation of the system.
 
The computation cost of our solution is dominated by the \psica algorithm, which itself is dominated by (1) token transforms (exponentiation) and (2) keyword PIR~\cite{SP:ACLS18, EPRINT:ALPRSSY19}. We'll first propose the parameters of our estimation in the following subsection, then analyze the performance of PIR in single-server operation, PIR in two-server operation, token transforms, and finally summarize the overall system performance.
  
\subsection{Parameters and token storage}
\label{subs:db-params}

Assuming that a contact token is generated every 15 minutes for approximately 20 hours a day, then each user sends 80 distinct 128-bit tokens per day. If we assume that a user also receives approximately the same number of tokens and the infectious period is 14 days for COVID-19, then each client receives a total of $n=1120$ over 14 days. If there are 5,000 new cases per day, the server receives $N = 1120 \times 5000 = 5.6 \times 10^6$ new tokens per day.

In \dect, the server maintains a list of tokens from positive patients for the duration of the infectious window. When a user is diagnosed positive for the disease, they give all of their sent contact tokens for the infection window (or the seeds to generate them) to the server. Rather than storing these by day they were exchanged, it's both more efficient and improves privacy for the server to store them by the day the server received the tokens. This way clients can query only for new tokens that have arrived since they last checked, rather than querying against the entire set.

More concretely, imagine the client did a query yesterday with all of their received tokens, $\sT_{R,C}$, against the sent tokens on the server, set $\widehat{\sT}_S$, and found no tokens in common. Today the client has the received tokens set $\sT_{R,C} + \sT_{R,C}'$, where $\sT_{R,C}'$ is the set of tokens received since the last query was performed. Similarly, the server has set $\widehat{\sT}_S + \widehat{\sT}_S'$, where $\widehat{\sT}_S'$ is the set of tokens received by the server from users diagnosed positive since yesterday. The set $\widehat{\sT}_S$ cannot contain any tokens in $\sT_{R,C}'$, but set $\widehat{\sT}_S'$ can contain tokens from either $\sT_{R,C}$ or $\sT_{R,C}'$. Therefore, the client only needs to check if $\widehat{\sT}_S'$ has any intersection with $\sT_{R,C} + \sT_{R,C}'$. Thus, if the server keeps the database by day received, clients query with their entire received token set ($\sT_{R,C} + \sT_{R,C}'$) against the server set only for the most recent day ($\widehat{\sT}_S'$).

\paragraph{Client's token storage.}  Storing both sent and received tokens requires 35 KiB of storage on the client (this can be reduced to 18 KiB if sent tokens are generated with a PRG and only the seed needs to be stored).

\paragraph{Server's token storage.} Assuming the server needs to keep 15 days of tokens in case clients are offline, the total storage for tokens needed is 1.25 GiB.

\subsection{Implementation optimization: Database shape}
\label{subs:impl_optz}
The bottleneck for scaling \psica to serve a large dataset to a large number of users is PIR. In order to scale up PIR, we propose using a bucket system similar to the password checkup design in \cite{EPRINT:ALPRSSY19}. First, the database will be split into $n_{shards}$ shards (sometimes referred to as megabuckets). Transformed tokens will be grouped into buckets, each bucket holding the same number of tokens, with dummies added as needed. Rather than performing keyword PIR, normal PIR with a bucket address is used. Since tokens are expected to have a uniform distribution (both before and after transformation), tokens should be uniformly distributed across shards and buckets. As such, the bucket addresses can simply be the first $\log_2(n_{shards} n_{buckets})$ most significant bits of the transformed token itself. Alternatively, a fast hash of the transformed token into the number of bits needed can be used. Recall that each transformed token is truncated to 74 bits before being stored in the database. We use the top bits of the token to be the database index and bucket address, and only store the remaining bits.

For example, if there are 5.6 million tokens in the server's set, the database can be sharded into 8 sets each with approximately 700,000 tokens (again, assuming a uniform distribution of tokens). If each shard holds $2^{18}$ buckets, then each bucket will hold $\ceil{\frac{700,000}{2^{18}}} = 3$ transformed tokens, with dummies added as necessary to pad buckets to the required length. The first three most significant bits of the transformed token are used as the shard number, and the following 18 bits of the transformed token are then the bucket address.  Since each transformed token is stored as $74 - 3 - 18 = 53$ bits, each bucket will be 20 bytes.

\paragraph{Security.} While the server does learn the first $\log_2(n_{shards})$ bits of the tokens being queried by the client, we consider that because the tokens (both before and after transformation) should be uniformly distributed across the token space, and that the numbers of tokens being queried is much larger than the number of shards (1120 vs 8 in the example above), in most cases the client will be querying all of the shards, therefore the server learns very little. This can also be scaled automatically by the server. If very few tokens are received on a given day, then the number of shards can be decreased. If a much larger number of tokens is received on a particular day, then the number of shards can be increased automatically.

The client learns the values of the other transformed tokens in the bucket and its size. Since the tokens have been transformed and truncated, and the client cannot reverse the transformation, the client doesn't learn anything useful as a result.

\subsubsection{Single Server PIR Performance}
Using the parameters from the example above, let us estimate the performance of such a configuration. As described in Section~\ref{subs:db-params}, we assume the client has 1120 tokens that must be compared against the set of new tokens received by the server since the last query, and that the client completed a query yesterday. If previous days need to be checked, each day will take approximately the same time.

Based on the performance figures for multi-query PIR in~\cite[Figure~12]{SP:ACLS18}\footnote{PIR servers on H16 instances (16-core 3.6 GHz Intel Xeon E5-2667 and 112 GB RAM), and clients on F16s instances (16-core, 2.4 GHz Intel Xeon E5-2673 and 32 GB RAM)}, with a database of $2^{20}$ items of 288 bytes each, and a batch size of 256 queries, \pirqe\ and \pirextract\ require 4.92 ms and 2.70 ms per query item on the client respectively. The communication cost of \pirqe\ is 96 KiB. The Server's \pirexpand\ and \pirans\ take 0.12 s and 0.08 s respectively, amortized across the batch, with a one-time preprocessing of the database of 4.26 s. The communication cost of \pirans\ is 384 KiB.  Assuming all of the database parameters scale appropriately, we estimate the performance of single-server PIR as applied to \dect as follows:

\begin{itemize}
  \item \textbf{Server}
  \begin{itemize}
      \item \textbf{Pre-processing.} The server requires a one-time pre-processing data phase which takes 1.06 seconds per shard. The shards can likely be processed in parallel.

      \item \textbf{\pirexpand.} $1120 \times 0.12 s \times \frac{2^{18}}{2^{20}} =$ 33.6 seconds
      
      \item \textbf{\pirans.} $1120 \times 0.08 s \times \frac{2^{18}}{2^{20}} \times \frac{20}{288} =$ 1.6 seconds
  \end{itemize}
  \item \textbf{Client}
  \begin{itemize}
      \item \textbf{\pirqe.} $1120 \times 4.92 ms \times \frac{2^{18}}{2^{20}} =$ 1.4 seconds

      \item \textbf{Key Caching} If the keys from previous days are cached (on either the server or client), then the client only needs to query 80 new items every day, this produces a computation time of $80 \times 4.92 ms \times \frac{2^{18}}{2^{20}} =$ 98 milliseconds.

      \item \textbf{\pirextract.} $1120 \times 2.70 \times \frac{20}{288} = $ 210 milliseconds
  \end{itemize}
  
  \item \textbf{Communication.} The client sends to the server a query for 1120 items, each of $96 \times \frac{\log(2^{18})}{\log(2^{20})}=86.4$ KiB, for a total query size of 94.5 MiB. It then receives the same number of answers, each of $384 \times \frac{20}{288}$ = 26.7 KiB, for a total answer size of 29 MiB. The total communication cost is approximately 124 MiB
  .
  \item \textbf{Communication with caching.} If server can cache the queries sent for the previous day, then only 80 items need to be added to the query. The query size is thus reduced to 7 MiB, and the total communication cost is reduced to 36 MiB.
  
\end{itemize}

\subsubsection{2-Server PIR Performance}
\label{subs:2PIR}

If 2 independently operated servers are available such that there will not be collusion between them, then the 2-server PIR model may offer performance benefits. The best two-server PIR is currently based on the distributed point function~\cite{EC:BoyGilIsh15,CCS:BoyGilIsh16}, in which 
$\pirqe$ and  $\pirexpand$ require $2N$ AES calls, where $N$ is the size of the server dataset. We omit the cost of \pirans\ and \pirextract\ in this PIR construction since they contain only bitwise operations. Using hardware-accelerated AES-NI as the underlying PRF, both $\pirqe$ and  $\pirexpand$ take approximately 31.1 milliseconds for a database of 5.6 million entries\footnote{10 cycles per AES, thus 360 AES per microsecond using 3.6 Ghz machine}. If implemented using keyword PIR based on cuckoo hashing (approximately three times the cost of regular PIR), this would require approximately 104 s for $3\times 1120$ \pirqe\ on the client without catching. Note that client can store their (short) keys from \pirqe\ of previous days, thus the client only needs to do $3\times 80$ new \pirqe\ per day, which requires $7.42$ s. We require another 104 seconds for \pirexpand\ on the server. 

Similar to single-server PIR, this can be improved by sharding the database and using buckets. If we use a similar configuration to the single-server model presented in the previous subsection (8 database shards, each with $2^{18}$ buckets containing 3 transformed tokens), then the time for \pirexpand\ becomes $1120 \times 2 \times 2^{18} \times \frac{10}{3.6\times10^9} =1.63$ seconds on the server's side. The time for \pirqe\ with and without caching is $80 \times 2 \times 2^{18} \times \frac{10}{3.6\times10^9} = 0.11$ seconds and 1.63 seconds, respectively. 

Since the communication cost of the short keys is small and we consider any single server untrusted, we recommend caching the (short) keys on the client. Each key has $832$ bits, which requires $2 \times (1120-80) \times 832$ (bits) $= 210$ KiB of storage.


\paragraph{Communication.} The client needs to query both servers with 1120 (short) keys, each of $\kappa\log(N)=832$ bits, transmitting $2 \times 1120 \times 832$ (bits) $=227.5$ KiB. The client then receives 1120 answers from both servers, each of size $3 \times 53$ bits (since each bucket contains 3 tokens), receiving $2 \times 1120 \times 3 \times 53$ (bits) $= 43$ KiB. This produce a total network bandwidth of 288 KiB.

We consider caching of the query keys on the server impractical, as both servers would have to independently maintain a set of caches per user. This could lead to consistency or security problems for a relatively small gain in bandwidth. The client's query would be reduced from 227.5 KiB to 16 KiB, a savings of 211.5 KiB. For the rest of the discussion here we will consider that caches in 2-server PIR are kept on the client only.

 \subsection{Token Transforms}
 
The \psica algorithm transforms tokens through exponentiation as a way of blinding tokens. As a basis for estimation, we assume that a single exponentiation calculation using libsodium takes approximately 54 microseconds\footnote{ In~\cite[Table~2]{C:PRTY19} it was found that $2^{21}$ exponentiations took 1148.1 seconds in a single thread using the miracl library. Since libsodium is approximately $10\times$ faster, we estimate the time per exponentiation as $1148.1 s / 2^{21} / 10 = 54 us$. Experiments were done on a server with an Intel(R) Xeon(R) E5-2699 v3 2.30GHz CPU and 256 GB RAM}.
 
If the server reuses the exponent $k$ (step 1, Figure \ref{fig:unbalance_psica}) across queries and users, then it can precompute all of the values $H(x_i)^k$ and store the results in a table. Thus the server only needs to compute $N=5.6 \times 10^6$ exponentiations per day, which can be done offline and not during a query. From our benchmarks, this takes approximately 5 minutes. For each user, the online portion of the exponentiation computations takes approximately 60 milliseconds. 
This can be reduced to 4.3 milliseconds if values from previous days are cached.

Assuming the client caches previously transformed tokens, the client must compute 80 new exponentiations in the preproceessing phase and another 80 exponentiations in the online phase, which we estimate will take approximately 4.32 milliseconds for each phase.

In terms of security, reusing the value of $k$ means that a client can obtain the transformed value of a particular token by first asking for the transforms of a set of tokens that doesn't include the desired token, and then asking for the transforms of the same set with the addition of the desired token, and finding the difference between the sets. A client could then use that to determine if that particular token is in the database by inspecting the results of the PIR query. However, this is indistinguishable from a client crafting arbitrary queries to the entire protocol as described in Section~\ref{subs:security_discussion}. It is still not possible for a client to determine the value of $k$, nor does it assist a malicious user to determine  which token is the source of their infection any more efficiently than already possible in the design of the system.

\paragraph{Communication.} The client sends and receives 1120 group elements twice. Using Koblitz K-283 elliptic curves, each element is 283 bits, for a total of $2 \times 1120 \times 283$ bits $=$ 77.4 KiB.

If the client caches the transformed tokens   from previous days, the client sends and receives $2\times 80$ group elements which, for a total of 5.5 KiB.

\subsection{Caching and Security}

\begin{table}[]
	\centering
	\begin{tabular}{|l||r|r|r|r||r|r|r|r|r|r|}
		\hline
		& \multicolumn{4}{c||}{\textbf{1-Server}} & \multicolumn{4}{c|}{\textbf{2-Server}} \\
		& \multicolumn{2}{c|}{No Cache}
		& \multicolumn{2}{c||}{Cache}
		& \multicolumn{2}{c|}{No Cache }& \multicolumn{2}{c|}{ Cache} \\ 
		
		All values in ms & Client & Server & Client & Server & Client & Server & Client & Server \\ \hline
		Token Transforms & 60 & 60 & 4 & 4 & 60 & 60 & 4 & 4 \\ \hline
		\pirqe & 1,378 &  & 98 &  & 1,631 &  & 117 &  \\ \hline
		\pirexpand &  & 33,600 &  & 33,600 &  & 1,631 &  & 1,631 \\ \hline
		\pirans &  & 1,556 &  & 1,556 &  & neg &  & neg \\ \hline
		\pirextract & 210 &  & 210 &  & neg &  & neg &  \\ \hline
		\textbf{Total} & 1,648 & 35,216 & 313 & 35,160 & 1,692 & 1,692 & 121 & 1,635 \\ \hline
		
	\end{tabular}
	\caption{Estimated (online) running time for elements of \psica across different implementation options. The client's running time does not include the waiting time for server's response. All times in milliseconds. ``neg" indicates the cost of bitwise operations in the 2-server PIR construction~\cite{EC:BoyGilIsh15,CCS:BoyGilIsh16}
	}
	\label{tbl:comp-summary}
	
	\medskip
	
	\centering
	\begin{tabular}{|l||r|r||r|r|}
		\hline
		& \multicolumn{2}{c||}{\textbf{1-Server}}
		& \multicolumn{2}{c|}{\textbf{2-Server}} \\
		
		All values in KiB & No Cache & Cache & No Cache & Cache \\ \hline \hline
		Token Transforms & 77 & 6 & 77 & 6 \\ \hline
		Query & 96,768 & 6,912 & 228 & 228 \\ \hline		
		Answer & 29,867 & 29,867 & 43 & 43 \\ \hline		
		\textbf{Total} & 126,712 & 36,784 & 348 & 277 \\ \hline		
		
		
	\end{tabular}
	\caption{Network costs for \psica implementation options. All numbers in KiB. Note that for 2-Server PIR all caches are kept on the client, as keeping caches across two servers is considered impractical.}
	\label{tbl:comm-summary}
\end{table}

In several of the options discussed in this section, caching previous values has been presented. While this can produce a savings in computation and communication, it does present a potential leakage of information. Because randomized values are not regenerated, it becomes possible for the server (or an observer at the right location) to determine which values are the same and which have changed. As such, the server can deduce which of the transformed tokens were in yesterday's set, which are new, and which values have been removed. It still cannot deduce the value of the tokens themselves, but this allows the server to guess at how many tokens have been exchanged on a particular day. This could be more (or less) interactions with others, or longer (or shorter) interactions with the same people, or a combination of the two.

Note that the server always knows the total number of tokens being queried, and thus can always guess at changes from one day to the next, but this allows more precise estimation of token volume. If this leakage is important, then dummies can be added to ensure that all days have the same number of items queried, or a random number of dummies added to mask the true size of the query. Or caching can be dropped entirely if this risk is considered too great. For this reason we present both options.

\subsection{Overall PSI-CA Performance Estimates} 
The performance estimates for the overall \psica algorithm are shown in Table~\ref{tbl:comp-summary} for computation and Table~\ref{tbl:comm-summary} for communication, each showing the effect of the implementation options discussed in this section. The 2-server approach reduces both server computational load and produces a large savings in network bandwidth, but requires an independent party and thus may increase infrastructure costs.

We also believe that single-server PIR is feasible. If the query is done in the background without the user waiting on a response, then the query can be done in the cloud as a lower-priority batch processing job. Thus the approximately 35 seconds required to complete the query is a non-issue, and server resources can be scaled up to meet the number of users required. This was an intentional tradeoff for network efficiency. If the server does some caching of the query keys, then only 37 MiB of network traffic is needed.

Alternatively, the number of database shards can be increased. This does reveal slightly more information about the client's tokens to the server, but this may be acceptable depending on the number of tokens and performance requirements.

We believe that the \dect solution proposed is feasible in practice. The database parameters can be tuned for different tradeoffs of network and computation usage. This will be studied further at implementation to determine the optimal configuration. 

 \remove{
Compared to other system in Table~\ref{tbl:comp-complx}:
\begin{itemize}
    \item Baseline system (these designs make public notifications): Downloading $N$ new tokens of users diagnosed, each token of size 128 bits, requires 89.6MB.
    \item PACT\cite{chan2020pact}, client exchanges $2\times 80$ group elements at contact event, and receives $2N$ group elements from the server. Each group element has 256 bits. Therefore, the communication cost of PACT's client is 358.4MB. 
\end{itemize}
}

\appendix

\section*{Acknowledgments}
We thank Min Suk Kang, Ilya Sergey, Jun Han, Xiaoyuan Liu, and Duong Hieu Phan for helpful discussion.  This material is in part based upon work supported by the National Science Foundation(NSF) under Grant No. TWC-1518899, DARPA under Grant No. N66001-15-C-4066, Center for Long-Term Cybersecurity (CLTC), and IC3 industry partners. Any opinions, findings, and conclusions or recommendations expressed in this material are those of the author(s) and do not necessarily reflect the views of NSF, DARPA, CLTC or IC3.

\bibliographystyle{plain}


\end{document}